\documentclass[12pt]{article}
\usepackage{latexsym,colordvi}

\usepackage{graphicx}
\newcommand{\tr}{^{t r}}
\newcommand{\Tr}{{\rm tr}}
\newcommand{\G}{{\cal G}}
\newcommand{\U}{{\rm U}}
\newcommand{\SU}{{\rm SU}}
\newcommand{\be}{\begin{equation}}
\newcommand{\ee}{\end{equation}}
\newcommand{\MeV}{{\rm MeV}}
\newcommand{\lt}{\left}
\newcommand{\rt}{\right}
\def\dslash{\partial\hspace{-.2cm}/}

\def\lesim{${\lower 2pt\hbox{$\scriptstyle
<$}\atop\raise 4pt\hbox{$\scriptstyle\sim$}}$} 
\def\grsim{${\lower2pt\hbox{$\scriptstyle >$} \atop\raise4pt\hbox 
{$\scriptstyle\sim$}}$} 

\begin{document}
\begin{center}
\begin{flushright}
     SWAT/02/351\\
September 2002
\end{flushright}
\vskip 10mm
{\LARGE
Evidence for BCS Diquark Condensation in the $3+1d$ Lattice NJL Model
}
\vskip 0.3 cm
{\bf Simon Hands and David N. Walters}
\vskip 0.3 cm
{\em Department of Physics, University of Wales Swansea,\\
Singleton Park, Swansea SA2 8PP, U.K.}
\vskip 0.3 cm
\end{center}
\noindent{\narrower
{\bf Abstract:} We present results of numerical simulations of the 3+1$d$
Nambu -- Jona-Lasinio model with a non-zero
baryon chemical potential $\mu$, with particular emphasis on the superfluid
diquark condensate and associated susceptibilities. The results, when
extrapolated to
the zero diquark source limit, are consistent with the existence of a 
non-zero BCS
condensate at high baryon density. The nature of the infinite volume and zero
temperature limits are discussed.
}
\section{Introduction}

Colour superconductivity (CSC) at low temperature and high baryon number
density is one of the most intensively studied topics in QCD thermodynamics
(for recent reviews see \cite{CSC}). The basic idea is that diquark pairs 
at the Fermi surface in quark matter
become bound and condense in a relativistic analogue
of the BCS mechanism for electronic superconductivity \cite{BCS}. Because of 
the strong attractive $q q$ force in QCD, however, estimates of the energy gap 
$\Delta$, which opens up at the Fermi surface, can be as large as 100MeV
\cite{BR}, with potentially important consequences for the physics of
compact stars. For instance, one intriguing possibility 
is that pairing between quarks
with differing Fermi momenta may lead to a crystalline phase in which 
translational invariance of the ground state is spontaneously broken
\cite{ABR}.

Analytic approaches to the problem, however, are restricted either to 
asymptotically high (and phenomenologically irrelevant) 
densities where perturbative QCD is applicable, or to 
self-consistent treatments of model theories which capture some but not all 
relevant physics. A first principles calculation using 
lattice QCD remains elusive due to the difficulties of performing Monte Carlo
simulations with baryon chemical potential $\mu\not=0$. There are, however,
simpler models which are simulable. Two Colour
QCD is a confining theory in which the lightest baryons
are bosonic $q q$ states. At high density its ground state has a gauge invariant
condensate $\langle q q\rangle\not=0$ which spontaneously breaks baryon number
symmetry leading to superfluidity. Because of the absence of a Fermi surface,
however, the phenomenon far more closely resembles a standard Bose-Einstein 
condensation, and is well-described by chiral perturbation theory \cite{KSTVZ};
this picture is confirmed by lattice simulation \cite{HMSMSO,TCQCD}.
The NJL model \cite{NJL}, by contrast, is a strongly-interacting 
model with fermionic excitations but no confinement. It can be treated by 
self-consistent analytic methods and has been applied
to CSC \cite{BR} as well as more general issues in low-energy QCD \cite{SPK}.

The NJL model with $\mu\not=0$ has also 
been simulated, on a 2+1 dimensional lattice. 
Although evidence for $q q$ pairing has been found in the 
scalar isoscalar channel \cite{HM}, BCS condensation does not take place
\cite{NJL3}, the argument being that long-wavelength fluctuations in the
phase of the condensate wavefunction wipe out long range order. Because
phase coherence remains, however, superfluidity
is realised in an unconventional ``thin film'' fashion. This picture, 
related to the low dimensionality of the system, is essentially
non-perturbative and would not be exposed by use of self-consistent methods.
It points to the potential importance of fully non-perturbative treatments
of even simple models in attempts to understand CSC. In this Letter we extend 
the lattice analysis of \cite{NJL3} to the NJL model in 3+1$d$, motivated by
{\em (i)} the possibility for a genuine BCS condensation in this case, and
{\em(ii)\/} the model's phenomenological relevance. In the following sections 
we review the lattice formulation of NJL, show how lattice parameters may be
chosen to match low energy QCD, and then report results from simulations with
$\mu\not=0$, paying particular attention to diquark
observables. We will
show that the high density phase is qualitatively different to that observed in
2+1$d$ and far more closely resembles a conventional BCS superfluid; however,
some issues concerning the 
thermodynamic limit need to be resolved before this conclusion
can become definitive.

\section{Lattice Model and Parameter Choice}

The model studied here is the 3+1$d$ version of that studied   
in \cite{NJL3}. In particular it has the action
\begin{equation}
S={\bf \Psi}\tr {\cal A} {\bf \Psi} 
	+\frac{2}{g^2}\sum_{\tilde{x}}\lt(\sigma^2 +
	\vec{\pi}.\vec{\pi}\rt),
\label{eq:action}
\end{equation}
where $g^2$ is the four-fermi coupling constant,
the bispinor ${\bf \Psi}$ is
written in terms of staggered isospinor fermionic fields via
${\bf \Psi}\tr =(\overline{\chi},\chi\tr)$, and the auxiliary bosonic
fields $\sigma$ and $\vec{\pi}$, which live on sites $\tilde{x}$ of a dual
lattice, are introduced in the standard way.
 Written in the Gor'kov basis the fermion matrix is
\begin{equation}
{\cal A}=\frac{1}{2}\lt(
\begin{array}{cc}
\overline{\jmath}\tau_2 & M \\ -M\tr & j\tau_2
\end{array}
\rt),
\end{equation}
where the matrix $M$ is defined by
\begin{eqnarray}
M^{p q}_{x y} & = & \frac{1}{2}\delta^{p q} 
\lt[\lt(e^\mu\delta_{y x+\hat{0}}-e^{-\mu}\delta_{y x-\hat{0}}\rt)
+ \sum_{\nu=1}^3 \eta_\nu(x)\lt(\delta_{y x+\hat{\nu}} -\delta_{y
x-\hat{\nu}}\rt) +2m_0\delta_{x y}\rt] \nonumber \\ 
& + &  \frac{1}{16}\delta_{x y}\sum_{\lt<\tilde{x},x\rt>}
	\lt(\sigma(\tilde{x})\delta^{p q} + i\epsilon(x)\vec{\pi}
(\tilde{x}).\vec{\tau}^{p q}\rt),
\end{eqnarray}
$m_0$ is the bare quark mass
 and the symbols $\eta_\nu(x)$, $\epsilon(x)$ are the
phases $(-1)^{x_0+\dots+x_{\nu-1}}$ and $(-1)^{x_0+x_1+x_2+x_3}$
respectively. The Pauli matrices $\vec{\tau}$ are normalised such that
${\rm t r}(\tau_i\tau_j)=2\delta_{i j}$, and $\lt<\tilde{x},x\rt>$
represents the set of 16 dual lattice sites neighbouring $x$.  
The diquark sources $j$ and $\overline{\jmath}$ are introduced to
allow us to measure the diquark condensate
 and play the same
${\rm r\hat{o}l e}$ as a bare quark mass in the lattice estimation of the chiral
condensate. (NB. The sources used are greater than those in \cite{NJL3}
by a factor of 2, which allows us to identify them with Majorana fermion
masses.) We simulate the action (\ref{eq:action}) using a hybrid Monte Carlo 
(HMC) 
algorithm which uses a functional weight $\mbox{det}{\cal AA^\dagger}$, thus 
doubling the number of fermion degrees of freedom, but which has the advantage
of of being exact. We treat the diquark terms in the
partially quenched approximation, ie. the sources $j$ and
$\overline{\jmath}$ are set to zero during the HMC update of the
bosonic fields, but are made non-zero during the measurement of
diquark observables. Field theoretically, this means that the formation of
virtual diquark pairs in the vacuum is suppressed.
This approach has the advantage that one can
examine many source strengths for only one (computationally
expensive) chain of field evolutions. 

Our formulation employs staggered
quarks, which, although preserving some of the model's continuum
chiral symmetry, increase the number of quark
degrees of freedom by a factor $N_c=4$. 
As the NJL model has no gauge degrees of freedom
we interpret these doublers phenomenologically
as ``colours''. The  model is therefore an effective $N_f=2$,
$N_c=4$ QCD-like theory. It should be stressed, however, 
that a diquark condensate in this model breaks global rather than local
symmetries, hence physically giving rise to superfluid rather than
superconducting behaviour. 
The continuum limit of the model yields 
the Lagrangian density \cite{HK,HMSMSO}
\begin{eqnarray}
{\cal L}_{{\rm cont}}&=&\overline{\psi}_i\lt(\dslash+m_o+\mu\gamma_0\rt)\psi_i
-\frac{g^2}{2}\lt[\lt(\overline{\psi}_i\psi_i\rt)^2 -
\lt(\overline{\psi}_i \gamma_5\otimes\vec{\tau}\otimes\gamma_5\psi_i\rt)^2\rt] 
\nonumber
\\ & &
+\frac{1}{2}\lt[j\lt({\psi_i}\tr C\gamma_5\otimes\tau_2\otimes  C\gamma_5
\psi_j\rt) + \overline{\jmath}\lt(\overline{\psi}_i
C\gamma_5\otimes\tau_2\otimes C\gamma_5{\overline{\psi}_j}\tr\rt)\rt],
\end{eqnarray}
which in the limits $m_0\rightarrow 0$ and $j,\bar\jmath\to0$ recovers the
$\SU(N_f)_L\otimes\SU(N_f)_R$ chiral and $\U(1)_B$ 
baryon number symmetries of QCD respectively\footnote{Doubling due to
$\mbox{det}{\cal AA^\dagger}$ results in an additional
$\SU(2N_c)$ global symmetry in the continuum limit.}.
In the tensor products the first matrix acts on spinor,
the second on isopinor, and the third on ``colour'' indices.
 
Due to the point-like nature of the four-fermi interaction, the NJL
model has no interacting continuum limit for $d\ge 4 $, leaving physics
dependent on the UV regularisation employed \cite{SPK,HK}. 
In order to ensure we simulate in a
physically plausible regime, we employ the methods used in \cite{SPK} to match
our
model's parameters to low energy, vacuum phenomenology, taking
advantage of the fact that a perturbative expansion in $1/N_c$ is possible in
four-fermi theories. We calculate quantities analytically to leading
order in $1/N_c$, the Hartree approximation,  using
staggered quark propagators defined on a $L_s^3\times L_t/2^4$
Euclidean blocked lattice \cite{KS},
with periodic boundary conditions in spatial dimensions and
anti-periodic boundary conditions in the temporal dimension. The momentum
loop integrals are discretised using mode sums and taken to the limit
$V^{-1}\rightarrow 0$ where $V=L_s^3L_t$. 

In particular
we calculate the dimensionless ratio between the pion decay rate
$f_\pi$ and the constituent quark mass $m^*=\Sigma+m_0$, where
$\Sigma\equiv\langle\sigma\rangle$. 
\be
\frac{f_\pi}{m^*}
=\sqrt{2N_c N_f}
\int^{\frac{\pi}{2}}_{-\frac{\pi}{2}}{\frac{d^4p}{\lt(2\pi\rt)^4}
\frac{\cos{2p_\nu}}{\lt[\tilde{p}^2+\lt(am^*\rt)^2
\rt]^2}}
\sqrt{\int^{\frac{\pi}{2}}_{-\frac{\pi}{2}}{\frac{d^4p}{\lt(2\pi\rt)^4}
\frac{1}{\lt[\tilde{p}^2+\lt(am^*\rt)^2\rt]^2}}},
\label{fpi/m}
\ee 
where $\tilde{p}^2=\sum_{\nu=0}^3{\sin^2p_\nu}$.

Fixing $f_\pi$ to its experimental value of 93MeV and $m^*$ to a
physically reasonable 350MeV, we
extract a dimensionless quark mass of $am^*=0.3251$, where $a$ is
the spacing between lattice points, which in this section we will
not set to unity. Calculating the product of 
$\Sigma$ and the dimensionless inverse coupling $\beta=a^2/g^2$,
\be
\Sigma\beta a=2N_c N_f am^*\int^{\frac{\pi}{2}}
_{-\frac{\pi}{2}}
\frac{d^4p}{\lt(2\pi\rt)^4}\frac{1}{\tilde{p}^2
+\lt(am^*\rt)^2}=0.1826,
\label{gapeqn}
\ee
allows us to derive that
$\beta=0.1826/{\lt(0.3251-am_0\rt)}$. 

Finally, calculating the mass of the pion $m_\pi$ allows us to fix
$m_0$, and therefore $\beta$. In the Hartree approximation the pion
propagator can be written as a series of connected bubble diagrams
\be
\frac{a^2}{\tilde{k}^2+{\lt(am_\pi\rt)}^2}=\frac{g^2}{2} 
- \frac{g^2\Pi_{\rm{p s}}(\tilde{k}^2)g^2}{4}
+\frac{g^2\Pi_{\rm{p s}}(\tilde{k}^2)g^2\Pi_{\rm{p s}}(\tilde{k^2})g^2}{8} 
- \ldots
=\frac{g^2}{2+g^2\Pi_{\rm{p s}}(\tilde{k}^2)}
\label{piprop}
\ee
in terms of the vacuum polarisation of the pion
\be
\Pi_{\rm{p s}}\lt(\tilde{k}^2\rt)=-2
\lt(\frac{m^*-m_0}{m^*g^2}\rt) \\
+2N_c N_f{\tilde{k}^2\over a^2}I\lt(\tilde{k}^2\rt) ,
\label{PIps simple}
\ee
where 
\be
I\lt(\tilde{k}^2\rt)=
\int^{\frac{\pi}{2}}_{-\frac{\pi}{2}}
{\frac{d^4p}{\lt(2\pi\rt)^4}
\frac{1} {\lt[\sum_\mu{\sin^2{\lt(
p+\frac{k}{2}\rt)_\mu}}+{\lt(am^*\rt)}^2 \rt]
\lt[\sum_\nu{\sin^2{\lt(p-\frac{k}{2}\rt)_\nu}}+{\lt(am^*\rt)}^2
\rt]}},
\label{I}
\ee
$p_\mu$ is the loop momentum shifted to be independent of $a$, and
$k_\mu/a$ is the physical momentum of the pion. 
The fact that there is a pole in (\ref{piprop}) at
$\tilde{k}^2=-a^2m_\pi^2$ combined with (\ref{I}) allows us to write
\be
m_\pi^2 = \frac{m_0}{m^*}\lt. \frac{1}{N_c N_f g^2 I\lt(\tilde{k}^2
\rt)} \rt|_{\tilde{k}^2=-a^2m_\pi^2}.
\ee

A bare quark mass of $am_0=0.002$ leads to a phenomenologically
acceptable pion mass of 138MeV and implies an inverse coupling
$\beta=0.565$. Finally, as we know both $am^*$ and
$m^*$ we can extract the lattice spacing in the large-$N_c$ approximation,
with the result $a^{-1}=1076\MeV $, or
$a\sim 0.2{\rm f m}$.

\section{Equation of State}

As has been previously discussed, in the limit $m_0\rightarrow 0$ the
NJL model has an exact $\SU(2)_L\otimes\SU(2)_R$ chiral symmetry,
which for sufficiently large $g^2$ is 
spontaneously broken in the vacuum to $\SU(2)_{{\rm isospin}}$, leading
to a dynamically generated quark mass $m^*=\Sigma$, and 3 massless
Goldstone pions. In the presence of a baryon chemical potential $\mu$ the
symmetry is approximately restored as $\mu$ is increased through some
onset scale $\mu_o\sim\Sigma$, with the
order of the transition being highly sensitive to the parameters employed
\cite{SPK}. 

We determine the nature of this transition in our physically
reasonable regime by studying the order parameter, the chiral condensate
$\left<\overline{\chi}\chi\right>$,
defined by
\be
\left<\overline{\chi}\chi\right>= 
	\frac{1}{V}\frac{\partial\ln{\cal Z}}{\partial m_{_0}},
\label{pbp}
\ee
where ${\cal Z}$ is the partition function. We also study the baryon
number density $n_B$;
\be
n_B = \frac{1}{2V}\frac{\partial \ln {\cal Z}}{\partial \mu}.
\ee
These are calculated as functions of $\mu$, both in the large-$N_c$
limit and by numerical lattice simulation,
using a repeated stochastic estimator for the
diagonal elements of the inverse fermion matrix. We
use a HMC algorithm with $\beta=0.565$, $m_0=0.002$
on $L_s^3\times L_t$ lattices with
$L_s=L_t=12$, $16$ and $20$ for various values of $\mu$, as well as on
lattices with $L_s\neq L_t$ at $\mu=1.0$. 
Approximately 500
equilibrated configurations separated by HMC trajectories of mean length 1.0
were generated in each run, with
measurements carried out on every other configuration. Statistical
errors were calculated using a jackknife estimate.
\begin{figure}[ht]
\centering
\includegraphics[width=12cm]{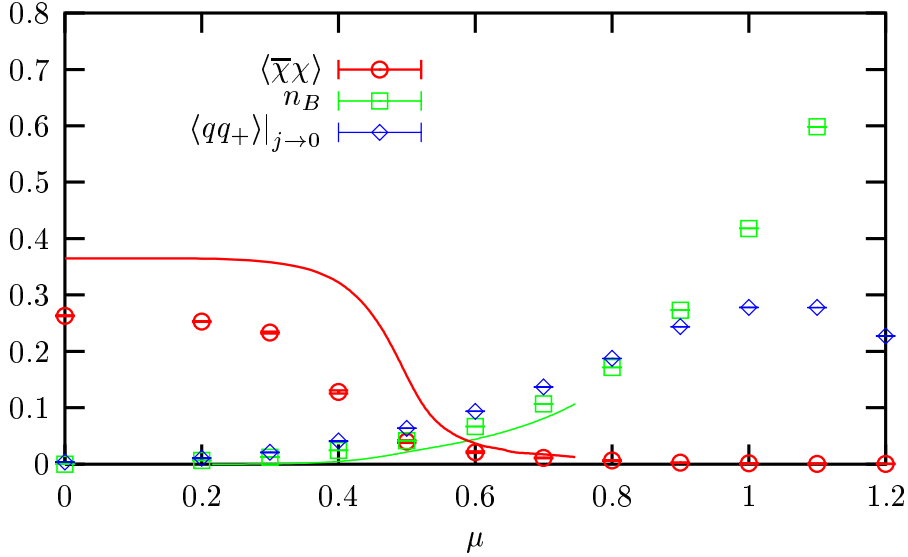}
\caption{Chiral condensate and number density as functions of $\mu$
extrapolated to $V^{-1}\rightarrow 0$ showing both the large-$N_c$
solution (solid curve) and lattice results (points). The diquark
condensate is plotted in the zero temperature, zero-source limit.}
\label{EofS}
\end{figure}

Our results are shown in Fig.~\ref{EofS}, where the
solid curve denotes the solution in the Hartree approximation, 
where $\langle\bar\chi\chi\rangle=2\Sigma\beta$, and the
points denote lattice results. We carried out a linear
extrapolation in $V^{-1}$ to the thermodynamic limit.  
To leading order in $1/N_c$ 
chiral symmetry is approximately restored via a crossover between
0.4\lesim$\mu$\lesim 0.6. The lattice data agree
qualitatively with this although 
both $\langle\bar\chi\chi\rangle$ 
and $\mu_o$ are about 30\% smaller, 
which we attribute to ${\cal O}(1/N_c)$ corrections.
$n_B$ increases approximately as a power of $\mu$.

\section{Diquark Condensation}

\subsection{Observables}

In order to explore the possibility of a
U(1)$_B$-violating BCS phase at high $\mu$ we
study both the diquark order parameter and susceptibilities \cite{NJL3}.
The operators
\be
q q_\pm(x)=\chi^{tr}\frac{\tau_2}{4}\chi (x) 
\pm\overline{\chi}\frac{\tau_2}{4}\overline{\chi}^{tr}(x)
\ee
allow us to define the diquark condensate as
\be
\left<q q_+\right>=\frac{1}{V}\frac{\partial\ln{\cal Z}}{\partial j_+},
\label{qq+}
\ee
where $j_{\pm}=j\pm\overline{\jmath}$. 
We also define the susceptibilities
\begin{eqnarray}
\chi_\pm =\sum_x\left<q q{_\pm}(0)q q_{_\pm}(x)\right>
&=&\hspace{-.2cm}\frac{1}{16}\sum_x
\lt<\chi\tr\tau_2\chi(0)\chi\tr\tau_2\chi(x)+\overline{\chi}\tau_2
\overline{\chi}\tr(0)\overline{\chi}\tau_2\overline{\chi}\tr(x)\rt>
\nonumber \\
& & \hspace{.56cm}\pm\lt<\chi\tr\tau_2\chi(0)\overline{\chi}\tau_2\overline{\chi}\tr(x)
+\overline{\chi}\tau_2\overline{\chi}\tr(0)\chi\tr\tau_2\chi(x)\rt>.
\label{susc}
\end{eqnarray}
These can be expressed as the sum of two connected contributions
corresponding to the two possible Wick contractions
\be
\chi=\lt[\lt<\lt(\Tr\Gamma\G_{x x}\rt)^2\rt>-\lt<\Tr\Gamma\G_{x x}\rt>^2\rt]
+\lt<\Tr\G_{0 x}\Gamma\G\tr_{0 x}\rt>\equiv\chi^s+\chi^{n s},
\ee
where $\G={\cal A}^{-1}$ is the Gor'kov propagator and $\Gamma$
projects out the appropriate components. By analogy with meson
physics we label these components ``singlet'' and ``non-singlet'' respectively. Both  $\lt<q q_+\rt>$ and $\chi^s_\pm$ are calculable
using the same stochastic estimation method used to measure
$\lt<\overline{\chi}\chi\rt>$ and $n_B$, whereas $\chi^{n s}_\pm$ is
measured using standard lattice methods for meson correlators. We fix
$j=\overline{\jmath}=j^*$ throughout. It is
interesting to note that although in most cases the singlet contributions
are found to be consistent with zero,
 in the low $\mu$ phase with large $j$, $\chi_+^s$ can
be up to $10-20\%$ the magnitude of $\chi_+^{n s}$. Therefore 
in contrast with the NJL model in 2+1$d$ \cite{NJL3} we
cannot ignore these contributions and assume that
$\chi_+\simeq\chi_+^{n s}$. 

From (\ref{susc}) it is straightforward to derive the Ward identity
\begin{equation}
\left.\chi_-\right|_{j_-=0}=\frac{\left<q q_+\right>}{j_+},
\end{equation}
which along with the ratio 
\begin{equation}
R=-\frac{\chi_{_+}}{\chi_{_-}}
\label{R}
\end{equation}
allows one to distinguish between the two phases as $j\to0$. 
With $\U(1)_B$ manifest, the two susceptibilities should be
identical up to a sign factor and $R$ should equal 1 in this limit.
If the symmetry is broken, the Ward identity
predicts that $\chi_-$ should diverge and $R$ vanish.

\subsection{Results}
\label{Results}

The susceptibilities were measured and $R$ calculated for the
aforementioned lattice sizes at various values of $\mu$. We
extrapolated the data using least-square linear fits to the limit
$L_t^{-1}\rightarrow 0$, which experience in 2+1$d$ \cite{NJL3} has shown
to be the best model of finite size corrections. 

\begin{figure}[ht]
\centering
\includegraphics[width=12cm]{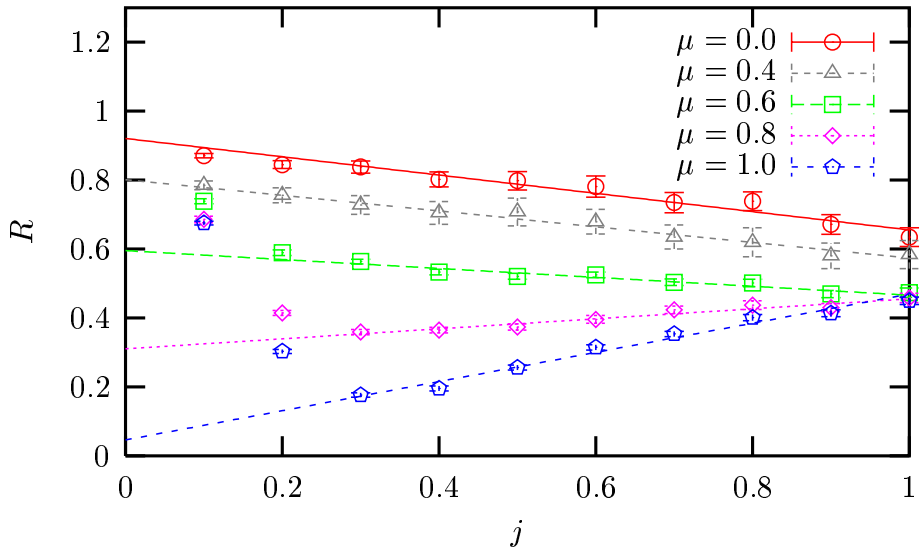}
\caption{$R$ as a function of $j$ for various $\mu$.}
\label{Rvsj}
\end{figure}

Fig.~\ref{Rvsj} shows this extrapolated data plotted against $j$ for
various values of $\mu$. One can immediately notice that although a
linear fit through the data for $j\geq0.3$
is reasonable, for $j<0.3$ the
data departs sharply from the fit,
especially
in the high density phase $\mu$\grsim0.6. 
Possible origins of this effect are explored in section \ref{FiniteV}.
Assuming that we can disregard the points with $j<0.3$, one can
see from Fig.~\ref{Rvsj} that for $\mu=0$ a linear fit is consistent
with a ratio of $R\approx 1$ corresponding to a manifest baryon
number symmetry as one would expect in the vacuum. As $\mu$ is
increased the value of the ratio decreases and as $\mu$ approaches one
inverse lattice spacing we see that $R\approx 0$, suggesting that
the $\U(1)_B$ symmetry is broken. 

\begin{figure}[ht]
\centering
\includegraphics[width=12cm]{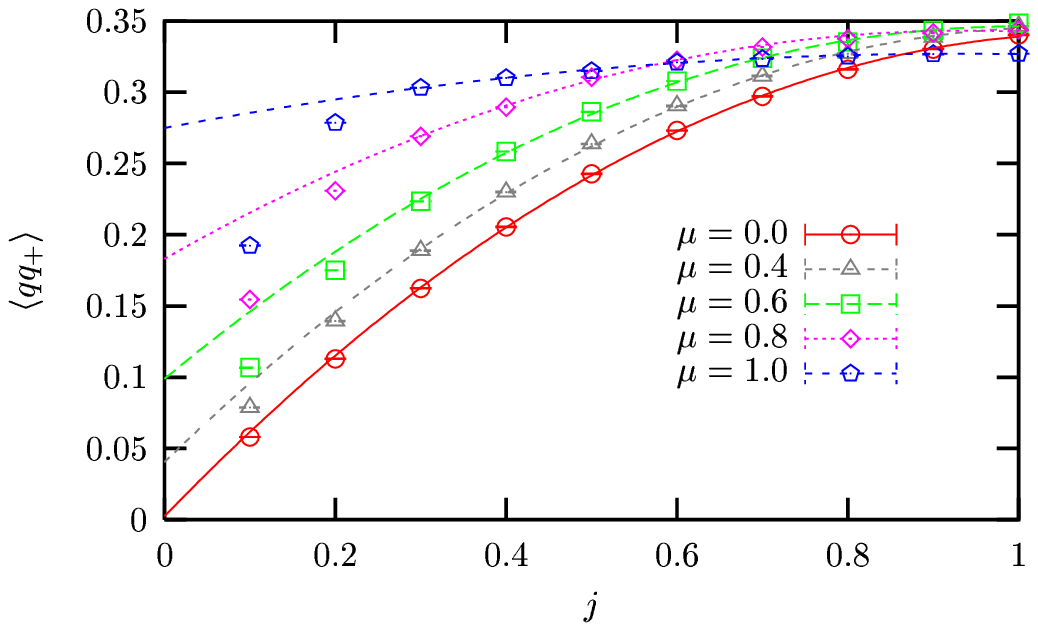}
\caption{$\lt<q q_+\rt>$ as a function of $j$ for various
$\mu$.}
\label{qqvsj}
\end{figure}

For more direct evidence of diquark condensation we study the diquark
condensate defined in (\ref{qq+}). Fig.~\ref{qqvsj} shows $\langle q q_+\rangle$
plotted against $j$ for various values of $\mu$. Again we
have extrapolated linearly to the limit $L_t^{-1}\rightarrow 0$. 
We use a quadratic fit through the data with $j\ge0.3$ but again one can
clearly see that for high $\mu$ the low $j$ points lie well below the
curves. Ignoring these points we extrapolate to $j\rightarrow
0$. For $\mu=0$ we find no diquark condensation as 
expected, but as $\mu$ increases from zero, so does $\lt<q q_+\rt>$.
We believe that the observations $\lim_{j\to0}R=0$ and
$\lim_{j\to0}\langle q q_+\rangle\not=0$ together
support the existence of a BCS phase at high chemical potential. 

Finally, $\langle q q_+\rangle$ is plotted as a function of $\mu$ in
Fig. \ref{EofS}. Although there is clearly a crossover from a phase
with no diquark condensation to one in which the diquark condensate
has a magnitude approximately that of the vacuum chiral condensate,
this crossover is far less pronounced than
in the chiral case. $\lt<q q_+\rt>$ increases approximately
as $\mu^2$, (ie. as the
surface area of the Fermi-sphere), but eventually saturates and even
decreases as $\mu$ increases through 1. This may be a cut-off artifact.
  The curvature
$\partial^2\lt<q q_+\rt>/\partial\mu^2$ is positive, in contrast to the
behaviour observed in simulations of Two Colour QCD in which there is 
no Fermi surface and U(1)$_B$ breaking proceeds via
Bose-Einstein condensation \cite{TCQCD}.
It is possible, of course, that this behaviour for intermediate
$\mu$
 is an
artifact of our poor control over the $j\to0$ extrapolation.

\subsection{The Low Temperature Lattice Fermi Surface}
\label{LowT}

We know, from experience in 2+1$d$,  that an
 extrapolation in $L_t^{-1}$ is the
 best description of of finite size corrections in this model.
 One natural progression would be
to simulate with a small $L_s$ and various $L_t$, saving both CPU time and
resources.  However, this approach presents a problem.

 As we
increase $\mu$ we build up a
Fermi surface, which in the continuum is a sphere smeared out by thermal
fluctuations over a region  $\delta k_F\sim 2T=(2L_t)^{-1}$.
In Euclidean space, an increase in $L_t$ corresponds directly to a
decrease in temperature, and at low temperatures, the Fermi-Dirac
distribution closely  resembles a step-function with all states with
$\sinh^{-1}\sqrt{\sum^3_{\nu =1}\sin^2k}\le E_F$ occupied, and all other states unoccupied. 
If we simulate with $L_t\gg L_s$, the smearing of the Fermi surface
will be too fine to be resolved on the coarse momentum lattice, and
as $\mu\approx E_F$ is increased the physics will be 
constant except when the Fermi sphere crosses each momentum mode,
turning any transition into a series of steps.

\begin{figure}[ht]
\centering
\includegraphics[width=12cm]{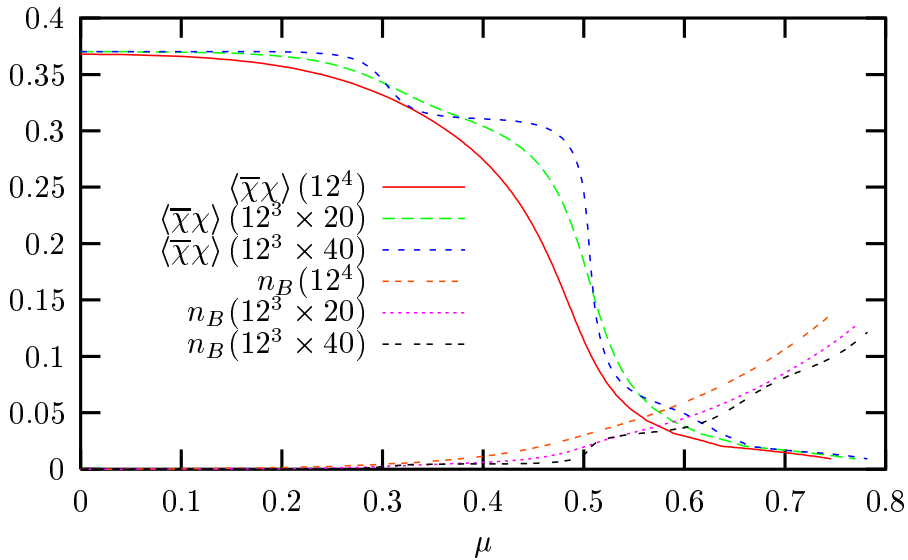}
\caption{Chiral condensate and number density in the large-$N_c$ limit
for various lattices.}
\label{Steps}
\end{figure}
Fig.~\ref{Steps} illustrates the effect that this has on the chiral
transition in the large-$N_c$ limit. $\lt<\overline{\chi}\chi\rt>$ and
$n_B$ are plotted on $12^4$, $12^3\times 20$ and $12^3\times 40$
lattices.  On the asymmetric lattices one can clearly see the discontinuities
caused as the Fermi surface crosses momentum modes. In the baryon
number density one can even observe a plateau as the first mode
becomes occupied. This behaviour was also observed in some initial
lattice simulations. In practice, it seems that the effect is only
prominent when $L_t$\grsim$3L_s/2$. 

\subsection{Finite Volume Effects}
\label{FiniteV}

\begin{figure}[ht]
\centering
\includegraphics[width=12cm]{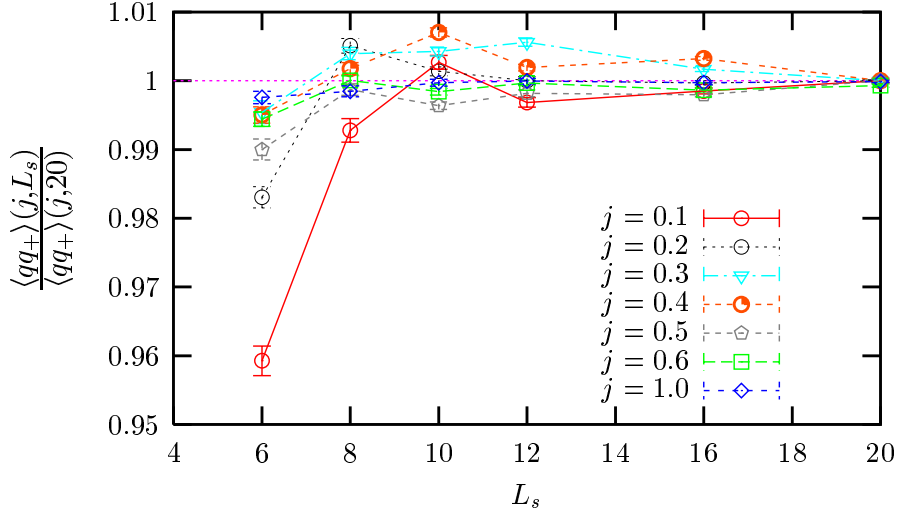}
\caption{$\lt<q q_+\rt>$ in the zero-temperature limit as a function
of the spatial extent of the lattice. The results are scaled such
that $\lt.\lt<q q_+\rt>\rt|_{L_s=20}\equiv 1$.}
\label{LsneqLt}
\end{figure}

To investigate the possibility that the low-$j$ discrepancies in
Figs.~\ref{Rvsj} \&~\ref{qqvsj} are due to finite volume errors,
it is necessary to
study spatial volume dependence in a controlled manner. 
We study all $L_s^3\times L_t$ lattices with 
$L_s=6$, 8, 10, 12, 16 \& 20 and $L_t=12$, 16 \& 20 for fixed
$\mu=1$. Being both well into the high-$\mu$
phase and far from the transition, any finite volume effects should be
easily identifiable.
We extrapolate to infinite temporal extent, corresponding physically
to zero temperature, using the fact that the dominant finite size
correction in our data is proportional to $L_t^{-1}$. However,   
Fig. \ref{LsneqLt} illustrates
that there is some residual $L_s$ dependence.   

$\lt<q q_+\rt>$ is plotted as a function
of spatial extent for various values of $j$, with data 
presented as fractions of $\lt.\lt<q q_+\rt>\rt|_{L_s=20}$ to enable
them to be displayed on the same axes. 
For $j\ge 0.5$, $L_s\ge 12$ the effect is negligible; the data are
consistent within errors. For $j<0.5$ we observe  
non-monotonic fluctuations, whose magnitude increases with
decreasing $j$.
A possible explanation is that with a small source, the region about
the Fermi surface in which diquark pairs would bind
may be too fine to be resolved on small 
spatial lattices.
Variational studies of the $N_f=2$, $N_c=3$ continuum NJL model at zero
temperature in a finite spatial volume \cite{PA.et.al} find that with no
diquark
source, a spatial extent of 7fm ($\sim 35$ lattice spacings) is
required before the model approximates its infinite volume limit. 
Below this volume, the BCS gap oscillates rapidly about its
thermodynamic limit solution. Although this may explain why, with a small
source, the condensate displays this non-monotonic $L_s$ dependence,
for $L_s\ge 12$ these fluctuations are less than a $1\%$ effect,
whereas the depletion of diquark pairs below the extrapolated curve in
Fig. \ref{qqvsj} is $\sim 30\%$.  
At this stage, therefore, we cannot claim to understand the origin of
the depletion of  diquarks at $j<0.3$.

\section{Summary and Outlook}

We have presented evidence for a BCS phase in the $3+1d$ lattice NJL
model at high chemical potential, by measuring a diquark order
parameter which appears, at $\mu=1$, to be of similar
magnitude to the chiral order parameter in the vacuum. This is
supported by a susceptibility ratio, $\lim_{j\to0}R(\mu=1)\approx 0$,
which suggests a broken
$\U(1)_B$ baryon number symmetry in the high-$\mu$ phase. Both features 
contrast sharply 
with the equivalent measurements in the 2+1$d$ NJL model \cite{NJL3}, 
where $\langle q q_+\rangle$ vanishes as a power of $j$ and $R$ is approximately
a constant independent of $j$, both indicative of a critical phase with
unbroken U(1)$_B$ symmetry at high density.

Our measurement of $\langle q q_+\rangle$ can be translated into an
order-of-magnitude estimate for the gap $\Delta$ using the following 
simple-minded
argument: the condensate is the number density of bound
diquark pairs, which in a 
BCS scenario is roughly equal to the volume of $k$-space within a distance
$\Delta$ of the Fermi surface divided by the density of states;
this yields $\langle q q_+\rangle\sim\Delta\mu^2$. Since 
at $\mu a=1.0$ we find $\langle q q_+\rangle a^3\sim m^*a\sim0.2 - 0.3$, 
it is quite plausible that the gap values of ${\cal O}$(100MeV) predicted in \cite{BR}
are possible.

Our claim to have observed a BCS phase in a lattice simulation is necessarily
provisional, however, because 
it depends on the discarding of data with
$j<0.3$, which in the high $\mu$ phase fall rapidly away from the
trends observed at higher $j$.
This threshold is very high when compared with chiral symmetry breaking at
$\mu=0$, where the bare quark mass
can be taken as low as $m_0=0.002$ with no adverse effects on
$\langle\bar\chi\chi\rangle$ on the volumes studied. It is also
larger by a factor of 3 than the equivalent threshold observed
in 2+1$d$ \cite{NJL3}.
The volume effects discussed in section \ref{FiniteV} do not resemble
those expected from the standard treatment of Goldstone fluctuations
\cite{Hasen}.
Variational studies in \cite{PA.et.al}
suggest that with a vanishing source, a $7$fm spatial box
($L_s\sim 35a$), which is currently unattainable, may be required for 
the BCS state to approximate its infinite volume limit. 

It is also possible that the low-$j$ 
discrepancy is a result of the partially quenched approximation, and
would be  resolved in a full unitary simulation.
Although a repeat of
the current study with a full simulation is currently beyond our
reach, were it deemed necessary,
it would be sufficient to repeat the study
presented in section \ref{FiniteV} with dynamical diquarks for a
limited range of $j$. It should be noted, however, that no significant effect on
diquark observables due to partial quenching was noted in \cite{NJL3}. 

We have chosen a value $m^*=350$MeV for the constituent quark mass in an
attempt to simulate with phenomenologically reasonable parameters.
The study of \cite{PA.et.al} was done with 
$m^*$ set to an equally reasonable
400MeV. Initial studies of the large-$N_c$ lattice model have determined
that with this choice, the chiral transition 
changes character to become much sharper and may even become first order
in the infinite volume limit.
It is an interesting coincidence that this change in
the order of the transition may occur in the ``physical''
region of parameter space. Not only is this worth exploring, but it
might make the determination of a BCS phase more clear cut. 
It is also desirable to determine the lattice spacing $a$ by direct measurement
of fermion and pion masses
so that we can convert our results into
physical units independent of the large-$N_c$ assumption.
Finally, we wish to study the fermion dispersion relation,
which will allow us to compare directly the chiral mass gap $\Sigma$ with the
BCS gap $\Delta$, these being, in principle, physically measurable
quantities.

\end{document}